\documentclass[aps,floats,amssymb,amsmath,prb,twocolumn,superscriptaddress,longbibliography]{revtex4-1}
\usepackage{graphicx}
\usepackage[english]{babel}
\usepackage{times}
\usepackage{color}

\begin{document}
\title{Magnetic impurities in spin-split superconductors}

\author{W.-V. van Gerven Oei}
\affiliation{Scientific Computing Laboratory, Center for the Study of Complex Systems, Institute of Physics Belgrade, University of Belgrade, Pregrevica 118, 11080 Belgrade, Serbia}

\author{D. Tanaskovi\'c}
\affiliation{Scientific Computing Laboratory, Center for the Study of Complex Systems, Institute of Physics Belgrade, University of Belgrade, Pregrevica 118, 11080 Belgrade, Serbia}

\author{R. {\v Z}itko}
\affiliation{Jo{\v z}ef Stefan Institute, Jamova 39, SI-1000 Ljubljana, Slovenia}

\begin{abstract}
Hybrid semiconductor-superconductor quantum dot devices are tunable
physical realizations of quantum impurity models for a magnetic
impurity in a superconducting host. The binding energy of the
localized sub-gap Shiba states is set by the gate voltages and
external magnetic field. In this work we discuss the effects of the
Zeeman spin splitting which is generically present both in the quantum
dot and in the (thin-film) superconductor. The unequal $g$-factors in
semiconductor and superconductor materials result in respective Zeeman
splittings of different magnitude. We consider both classical and
quantum impurities. In the first case we analytically study the
spectral function and the sub-gap states. The energy of bound states
depends on the spin-splitting of the Bogoliubov quasiparticle bands as
a simple rigid shift. For the case of collinear magnetization of
impurity and host, the Shiba resonance of a given spin polarization
remains unperturbed when it overlaps with the branch of the
quasiparticle excitations of the opposite spin polarization. In the
quantum case, we employ numerical renormalization group calculations
to study the effect of the Zeeman field for different values of the
$g$-factors of the impurity and of the superconductor. We find that in
general the critical magnetic field for the singlet-doublet transition
changes non-monotonically as a function of the superconducting gap,
demonstrating the existence of two different transition mechanisms:
Zeeman splitting of Shiba states or gap closure due to Zeeman
splitting of Bogoliubov states. We also study how in the presence of
spin-orbit coupling, modeled as an additional non-collinear component
of the magnetic field at the impurity site, the Shiba resonance
overlapping with the quasiparticle continuum of the opposite spin
gradually broadens and then merges with the continuum.
\end{abstract}

%\pacs{...}

\maketitle

\newcommand{\gbk}{g_\mathrm{bulk}}
\newcommand{\bbk}{b_\mathrm{bulk}}
\newcommand{\bimp}{b_\mathrm{imp}}
\newcommand{\expv}[1]{\left\langle #1 \right\rangle}
\newcommand{\vc}[1]{{\bf{#1}}}
\newcommand{\vck}{{\vc{k}}}
\newcommand{\gimp}{g_\mathrm{imp}}
\newcommand{\sgn}{\mathrm{sgn}}

\section{Introduction}

The interest in bound states induced by magnetic impurities in
superconductors, predicted in the early works of Yu, Shiba, and
Rusinov \cite{yu1965,shiba1968,rusinov1969}, has been recently revived
by the advances in the synthesis and characterization of
semiconductor-superconductor nanostructures
\cite{goffman2000,pillet2010,Deacon:2010jn,hybrid2010,bretheau2013,janvier2015}
and in the tunneling spectroscopy of magnetic adsorbates on
superconductor surfaces
\cite{yazdani1997,ji2008,franke2011,ruby2015,randeria2016,Hatter:2016kg}.
In particular, hybrid devices based on quantum dots can be used as
fully controllable physical realizations of quantum impurity models
with gapped conduction bands
\cite{sakurai1970,satori1992,sakai1993,salkola1997,flatte1997prl,flatte1997prb,yoshioka2000,morr2003,balatsky2006,bauer2007,rodero2011,dirks2011transport}.
The ground state of the quantum dot can be tuned to be either a spin
singlet or a spin doublet depending on the impurity level and the
hybridization with the bulk superconductor
\cite{buitelaar2002,pillet2010,Deacon:2010jn,maurand2012,martin2012andreev,jarrell1990gap}. The Coulomb
interaction on the quantum dot favors the spin doublet ground state,
while the spin singlet can be stabilized by the Kondo effect or by
pairing due to the superconducting proximity effect
\cite{vecino2003,oguri2004josephson,choi2004josephson,karrasch2008,meng2009}.
The position of the in-gap (Shiba) resonances, as determined from the
tunneling conductance, agrees even quantitatively with the
calculations based on the simple single-orbital Anderson impurity
model \cite{pillet2013,lee2016condmat}.

Very recently, research has focused on the effects of the magnetic
field on the in-gap states
\cite{mourik2012,lee2012,das2012,deng2012,chang2013,lee2014,zitko2015shiba,wentzell2016,jellinggaard2016,bujnowski2016andreev}
because systems of this class have been proposed as possible building
blocks for topologically ordered systems exhibiting Majorana edge
states \cite{kitaev2001,NadjPerge:2013et,NadjPerge2014,ruby2015majo}.
These are significant for fundamental reasons and might also find
application in quantum computation
\cite{sau2010,alicea2011,dassarma2015}. When an external magnetic
field is applied to a thin-film superconductor in the parallel
(in-plane) direction, the superconducting state persists to relatively
large fields. The quasiparticle states become, however, strongly spin
polarized and the coherence peaks in the density of states become
Zeeman split
\cite{meservey1970,tedrow:1971hp,Tedrow:1973fm,Meservey:1994bd,Eltschka:2014dn}:
systems in this regime are known as spin-split or Zeeman-split
superconductors, and play a key role in the emerging field of
superconducting spintronics \cite{linder2015}. The spectral function
of a spin-split superconductor has two band edges with diverging
coherence peaks separated by the bulk Zeeman energy, reflecting the
fact that the Bogoliubov excitations have spin-dependent energies
$E_{k\sigma}=\sqrt{\xi_k^2+\Delta^2}+\gbk \mu_B B \sigma$. Here
$\xi_k=\epsilon_k-\mu$ is the energy level $\epsilon_k$ of electron
with momentum $k$ measured with respect to the chemical potential
$\mu$, $\Delta$ is the gap, $\gbk$ is the $g$-factor of the
superconductor, $\mu_B$ is the Bohr magneton, $B$ is the magnetic
field, and $\sigma=\pm 1/2$ is the quasiparticle spin. Since the Shiba
states can be considered as bound states of Bogoliubov quasiparticles,
the spectral properties of magnetic impurities in spin-split
superconductors are modified.

The theoretical work has, so far, mainly focused on the effect of a
local magnetic field applied on the position of the impurity only
\cite{zitko2015shiba,wentzell2016}. For bulk electrons in the normal
state, this approximation is usually justified because the impurity
magnetic susceptibility is typically much larger ($\chi_\mathrm{imp}
\propto 1/T_K$, where $T_K$ is the Kondo temperature) than that of the
bulk electrons (Pauli susceptibility, $\chi_\mathrm{bulk} \propto \rho
\propto 1/D$, where $\rho$ is the density of states at the Fermi level
and $D$ is the bandwidth). In superconductors, however, the Zeeman
splitting of the Bogoliubov quasiparticle bands and the Zeeman
splitting of the doublet sub-gap states are of comparable magnitude:
the splitting of the first is simply the Zeeman energy $\gbk \mu_B B$,
while the splitting of the second is ${\tilde g}_\mathrm{imp} \mu_B
B$, where ${\tilde g}_\mathrm{imp}$ is the impurity $g$-factor
$g_\mathrm{imp}$ renormalized by the coupling with the bulk.
Generically, both splittings are comparable with the possible
exception of nanowire quantum dots made of materials with extremely
strong spin-orbit (SO) coupling and hence very high bare
$g_\mathrm{imp}$. For this reason, it is important to include the
Zeeman terms both in the impurity and in the bulk part of the
Hamiltonian.

We introduce the ratio $r$ of the Land\'e $g$-factors which describe
the magnitude of the Zeeman splittings:
\begin{equation}
r=\gbk/\gimp.
\end{equation}
For many elemental superconductors the $g$ factor is close to the free
electron value, $\gbk \approx 2$. In semiconductors the $g$ factor 
usually differs strongly from this value due to SO coupling. The
effective $g$ factors are quite variable \cite{csonka2008}: they can
be very large positive, as well as very large negative, or can even be
tuned close to 0. The control of $g$ can be achieved through strain
engineering \cite{nakaoka2005}, nanostructuring \cite{vanbree2016}, or
by electrical tuning in quantum dots
\cite{csonka2008,Schroer:2011ev,deacon2011,ares2013}. In the $r=0$ limit, the Zeeman
term is only present on the impurity site: this limit is appropriate
for materials with very large positive or negative $g$ factor, where
the Zeeman splitting in the superconductor is indeed negligible.
Another special limit is $r=1$, where all sites (bulk and impurity)
have the same $g$-factor. In general, however, the value of $r$ is
essentially unconstrained.

Here we study, using analytical calculations for a classical impurity
(with no internal dynamics) and with the numerical renormalization
group (NRG) method
\cite{wilson1975,krishna1980a,yoshioka2000,hofstetter2000,resolution,odesolv,hecht2008,bulla2008}
for a quantum impurity (which incorporates the effect of spin flips),
the spectral properties of the Shiba states.  In the classical case we
perform a calculation along the lines of
Refs.~\onlinecite{yu1965,shiba1968,rusinov1969}, but include the
effect of the Zeeman term in the superconductor. In the quantum case
we focus on the single-orbital Anderson impurity and discuss the
changes in the singlet-doublet phase transition as the ratio of the
$g$-factors of the impurity and the bulk is varied. We study the fate
of a sub-gap resonance when it approaches the continuum of the
Bogoliubov quasiparticles with the opposite spin, with and without the
additional transverse magnetic field that mimics non-collinearity in
the presence of SO coupling.

%Clogston-Chandrasekhar limit 

%FFLO state \cite{ff1964,lo1964}

\section{Classical impurity}

Initially, the impurity is described using a quantum mechanical
spin-$S$ operator, which is exchange coupled with the spin-density of
the conduction band electrons at the position of the impurity at
$\vc{r}=0$. The corresponding Hamiltonian is $H=H_\mathrm{BCS} +
H_\mathrm{imp}$ with
\begin{equation}
\begin{split}
H_\mathrm{BCS} &= \sum_{k\sigma} \xi_k c^\dag_{\vck,\sigma}
c_{\vck,\sigma} - \Delta \sum_k 
\left( c^\dag_{\vck,\uparrow} c^\dag_{-\vck,\downarrow} + \text{H.c.} \right) \\ 
&+ \sum_k \bbk s_{z,\vck},
\end{split}
\end{equation}
and
\begin{equation}
H_\mathrm{imp} = J \vc{S} \cdot \vc{s}(\vc{r}=0),
\end{equation}
where $\bbk=\gbk \mu_B B$ is the magnetic field expressed in the
energy units (i.e., the Zeeman splitting), $s_{z,\vck} =
\frac{1}{2}\left( n_{\uparrow,\vck} - n_{\downarrow,\vck} \right)$,
and $\vc{s}(\vc{r}=0) = \frac{1}{N} \sum_\vck s_{\vck}$.  $J$
is the exchange coupling between the impurity and the host. All other
quantities have already been defined in the previous section. The
classical impurity limit consists of taking the $S \to \infty$ limit
while keeping $JS=\text{const}$. In this limit, the longitudinal
component of the exchange interaction persists, while the transverse
(spin-flip) components decrease as $1/S$ and hence drop out of the
problem. The Hamiltonian then becomes non-interacting. We introduce
the effective local field 
\begin{equation}
h=JS
\end{equation}
and the dimensionless impurity coupling parameter 
\begin{equation}
\alpha=\pi \rho h/2 = \pi \rho J S/2,
\end{equation}
where $\rho$ is the density of states (DOS) at the Fermi level in the
normal state. We will first assume that the bulk field $\bbk$ and the
effective local field $h$ are collinear and of the same sign. To be
specific, we choose $\bbk>0$, $h>0$.

\begin{figure*}
\centering
\includegraphics[clip,width=\textwidth]{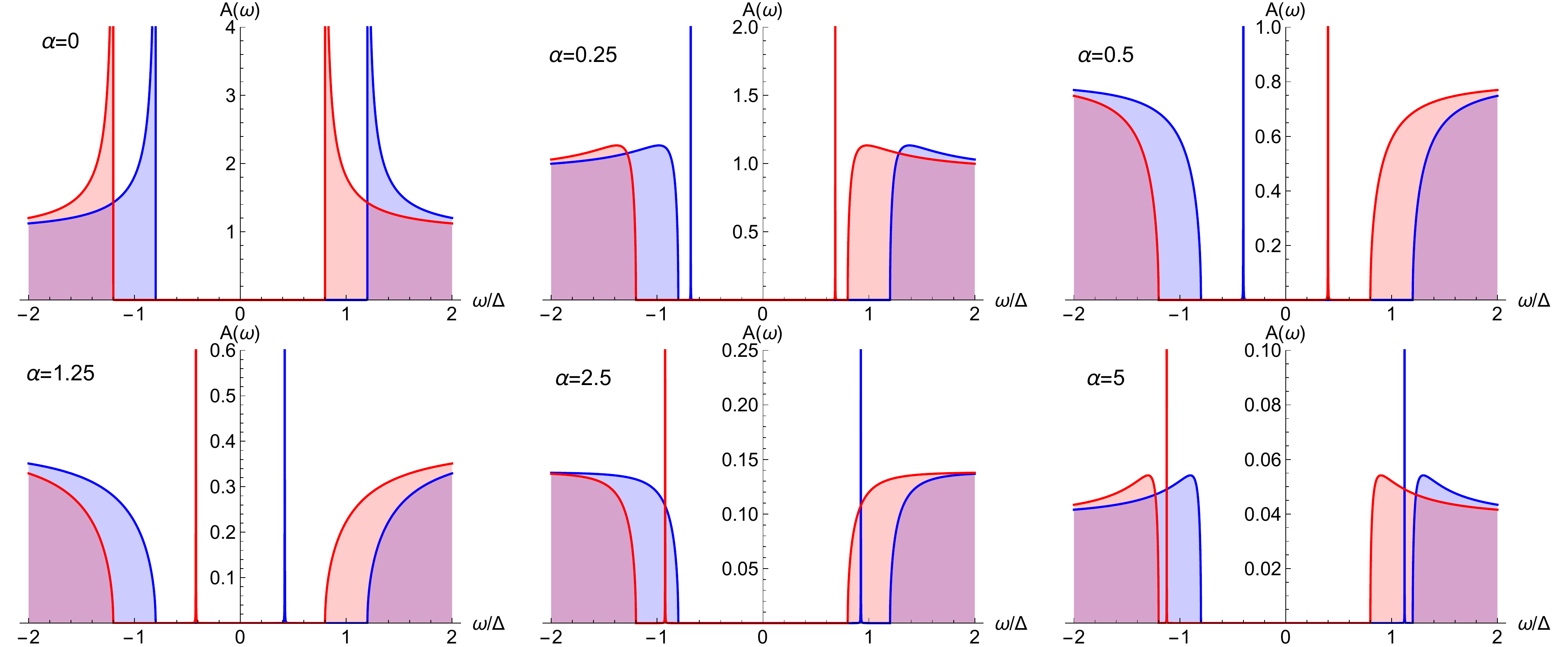}
\caption{Spin-projected spectral functions (blue for spin-up, red for
spin-down) for a range of the dimensionless impurity coupling
$\alpha=\pi\rho JS/2$ in a Zeeman-split superconductor
with $\bbk/\Delta=0.4$.}
\label{classical1}
\end{figure*}

The non-perturbed Green's function of the Zeeman-split superconductor
is
\begin{equation}
G^0_{k}(z) = \frac{(z-\bbk/2)\tau_0 + \epsilon_k \tau_3 -\Delta
\tau_1}{(z-\bbk/2)^2-(\epsilon_k^2+\Delta^2)}.
\end{equation}
Here $\tau_1,\tau_2,\tau_3$ are the Pauli matrices, $\tau_0$ is the
identity matrix, and $z$ is the frequency argument. To obtain the
local Green's function at the origin, $G^0_\mathrm{loc}$, we sum over
the momenta $k$ and switch over to an integral over energies assuming
a flat DOS in the normal state. In the wide-band limit we find
\begin{equation}
G^0(z) = - \pi \rho \frac{(z-\bbk/2)\tau_0-\Delta
\tau_1}{\sqrt{\Delta^2-(z-\bbk/2)^2}}.
\end{equation}
The Dyson's equation to include the impurity effect can be written
as \cite{yu1965,shiba1968,rusinov1969}
\begin{equation}
[G(z)]^{-1} = [G^0(z)]^{-1} - h \tau_0.
\end{equation}
We have
\begin{equation}
[G^0(z)]^{-1} = - \frac{\sqrt{\Delta^2-\left(z-\frac{\bbk}{2} \right)^2}}{\pi
\rho [\left(z-\frac{\bbk}{2}\right)^2-\Delta^2]} [(z-\bbk/2)\tau_0+\Delta\tau_1],
\end{equation}
and finally
\begin{equation}
G(z) = -\pi \rho \frac{1}{D} \begin{pmatrix}
a & \Delta \\ \Delta & a
\end{pmatrix},
\end{equation}
where
\begin{equation}
\begin{split}
D &= 2 \alpha \left(\frac{\bbk}{2}-z\right) + (\alpha^2-1)
\sqrt{\Delta^2-\left(\frac{\bbk}{2}-z\right)^2}, \\
a &= \bbk/2-z+\alpha\sqrt{\Delta^2-(\bbk/2-z)^2} .
\end{split}
\end{equation}
The spin-up spectral function is $A_\uparrow(\omega) =-(1/\pi) \Im
G_{11}(\omega+i\delta)$, while the spin-down spectral function is
$A_\downarrow(\omega) = -(1/\pi) \Im[ -G_{22}(-\omega-i\delta) ] = -(1/\pi)
\Im G_{22}(-\omega+i\delta)$.

The 11 (spin-up) matrix component of $G(z)$ has two poles:
\begin{equation}
\omega_{1,2} = \bbk/2 \pm \Delta \frac{1-\alpha^2}{1+\alpha^2}.
\end{equation}
Only one pole has a finite residue. For $h>0$ (hence $\alpha>0$) we
find a sub-gap resonance in the spin-up spectral function at
\begin{equation}
\label{oup}
\omega_\uparrow = \bbk/2 - \Delta \frac{1-\alpha^2}{1+\alpha^2}.
\end{equation}
Conversely, the spin-down spectral function has a resonance at
$\omega_\downarrow = -\omega_\uparrow$:
\begin{equation}
\label{odo}
\omega_\downarrow = -\bbk/2 + \Delta \frac{1-\alpha^2}{1+\alpha^2}.
\end{equation}
We emphasize that the spin-projected spectral functions have a single
sub-gap resonance, one for each spin. This is to be contrasted with
the behavior of the quantum model discussed in the following section
which has (in the spin-singlet regime for finite magnetic field) two
resonances in each spin-projected spectral function. This is a clear
indication of the different degeneracies of states in the classical
and quantum impurity models.

Some representative spectra are plotted in Fig.~\ref{classical1}.  The
$\alpha=0$ case corresponds to the limit of a clean Zeeman-split
superconductor. Each quasiparticle continuum branch has a
characteristic inverse square root divergence at its edge. 

For small $\alpha=0.25$, the Shiba bound states emerge out of the
quasiparticle continuum, the spin-up resonance in the negative part of
the spectrum, and the spin-down resonance in the positive part, in
line with Eqs.~\eqref{oup} and \eqref{odo} for small $\alpha$. The
shift by $\bbk/2$ is expected, since the spin-up Shiba state is
generated by the Bogoliubov states with spin up, which are themselves
shifted by the same amount. Conversely, the spin-down Shiba state is
generated as a linear superposition of Bogoliubov states with spin
down which are shifted by $-\bbk/2$. We observe that {\it all four}
branches of the quasiparticle band lose their inverse square-root
singularity and contribute spectral weight to the nascent Shiba state, see also Ref.~\onlinecite{bujnowski2016andreev},
not only the ``inner'' ones (spin-up occupied and spin-down
unoccupied). 

With increasing $\alpha$, the Shiba states move toward the
gap center (chemical potential) and they cross when the
condition
\begin{equation}
\bbk/2 = \Delta \frac{1-\alpha^2}{1+\alpha^2}
\end{equation}
is met, i.e., at
\begin{equation}
\alpha^* = \frac{\sqrt{1-\bbk/2\Delta}}{\sqrt{1+\bbk/2\Delta}}.
\end{equation}
For $\bbk/\Delta=0.4$, as used here, this happens at $\alpha^* \approx
0.82 < 1$. This signals the occurrence of the quantum phase transition
in which the fermion parity of the (sub)system changes. We also note
that alternatively, for constant $\alpha<1$, the transition can be
driven by the external magnetic field.

For still larger $\alpha=2.5$, the spin-up Shiba resonance overlaps with
the spin-down quasiparticle continuum (and vice versa for the
spin-down Shiba resonance), but since the spin is assumed to be a good
quantum number there is no broadening of the Shiba resonances. (See
below, Sec.~\ref{broad}, for a discussion of the SO effects in
the case of a quantum impurity.)

For very large values of $\alpha$, the Shiba states eventually merge
with the continuum again. This trend is accompanied by the
reappearance of the inverse square-root resonances, an indication of
which is visible for $\alpha=5$ in Fig.~\ref{classical1}.

We now discuss the case of anti-aligned fields, taking $\bbk>0$ and
$h<0$. In this case, for small $|\alpha|$ the spin-up Shiba state
occurs at
\begin{equation}
\omega_\uparrow = \bbk/2 + \Delta \frac{1-\alpha^2}{1+\alpha^2},
\end{equation}
and hence overlaps with the continuum of spin-down quasiparticles for
$|\alpha| < 1/\sqrt{2\Delta/\bbk-1}$. The quantum phase transition
occurs for
\begin{equation}
|\alpha^*| = \frac{\sqrt{1+\bbk/2\Delta}}{1-\bbk/2\Delta} > 1.
\end{equation}
For large $|\alpha|$ the Shiba states again merge with the continuum at
the inner edges of the Bogoliubov bands. The regimes that the system
goes through for $\alpha < 0$ are thus in the opposite order to those
for $\alpha > 0$.

The main deficiency of the impurity model in the classical limit is
the reduced multiplicity of the sub-gap states. Physically, this is
due to the fact that in the classical limit the effective impurity
potential for particle-like excitations is attractive for one spin
orientation and repulsive for the other, hence a single bound state is
generated for a given spin orientation. The spin-flip processes in the
quantum model lead to a situation where the effective potential is
attractive for both spin polarizations, hence twice the degeneracy. We
discuss this more general situation in the following section.

\section{Quantum impurity}

\subsection{Model and method}

We consider a single spin-$\frac{1}{2}$ impurity level with
on-site Coulomb interaction. The Hamiltonian is given by
\begin{align}\label{model}
H & = \sum_{{\bf k},\sigma}\epsilon_{\bf k} c^\dagger_{{\bf k}\sigma}c_{{\bf k}\sigma} - \Delta\sum_{\bf k}( c^\dagger_{{\bf k}\uparrow}c^\dagger_{-{\bf k}\downarrow}+\text{H.c.}) \nonumber
\\
& + V\sum_{{\bf k},\sigma}(d^\dagger_{\sigma}c_{{\bf k}\sigma}+\text{H.c.}) + \epsilon_d \sum_{\sigma} n_\sigma + Un_\uparrow n_\downarrow \nonumber
\\
&+\gimp \mu_B (BS_z + B_x S_x) + \gbk\mu_B B \sum_{\bf k}s_{z,{\bf k}} .
\end{align}
$d_\sigma^\dagger$ is the creation operator on the impurity which is
hybridized with the bulk by $V$ and has the energy level $\epsilon_d$.
$n_\sigma = d_\sigma^\dagger d_\sigma$, $S_z=\frac{1}{2}(d^\dagger_\uparrow d_\uparrow -
d^\dagger_\downarrow d_\downarrow)$, $S_x=\frac{1}{2} (d^\dag_\uparrow
d_\downarrow+d^\dag_\downarrow d_\uparrow)$, $s_{z,{\bf
k}}=\frac{1}{2}(c^\dagger_{{\bf k}\uparrow}c_{{\bf k}\uparrow} -
c^\dagger_{{\bf k}\downarrow}c_{{\bf k}\downarrow})$. The magnetic
field $B$ couples with the quantum dot by the $g$-factor equal to
$\gimp$ and with the superconductor by $\gbk$. The transverse magnetic
field which can flip the spin is introduced through the parameter
$B_x$. We will consider a flat particle-hole symmetric band of
half-width $D$ so that $\rho=1/2D$.  The hybridization strength is
characterized by $\Gamma = \pi \rho V^2$.
%We use the energy units of the half bandwidth $D$.

We employ the NRG method to solve the problem. There are two ways to
introduce a bulk Zeeman field in the NRG: as local Zeeman terms on all
sites of the Wilson chain, or through a separate discretization of
spin-up and spin-down densities of states shifted by the Zeeman term
\cite{hock2013}. The former approach is suitable for models with a
spectral gap, as discussed here, while the latter has to be used for
spin-polarized metals with finite DOS at the Fermi level. We use a
fine discretization mesh with twist averaging over $N_z = 64$ grids so
that high spectral resolution is possible inside the gap and in the
vicinity of the gap edges, which are the regions of main interest in
this work. The only conserved quantum number in the presence of an
external field along the $z$-axis is the projection of total spin
$S_z$, i.e., the problem has U(1) spin symmetry. Other parameters are
$\Lambda = 2$, the NRG truncation cut-off energy is $10 \epsilon_N$
where $\epsilon_N \propto \Lambda^{-N/2}$ is the energy scale at the
$N$-th step of the iteration and at least 200 states were used at late
iterations $N$ when the gap is opened. The spectral functions are
computed with the DMNRG algorithm \cite{hofstetter2000} with the $N/N
+ 2$ scheme for patching the spectral functions. This approach allows
maximal spectral resolution at zero temperature. The broadening is
performed on a logarithmic mesh with a small ratio $r = 1.01$ between
two energies outside the gap, and on a linear mesh inside the gap. As
can be seen in the figures further down, the use of these different
broadening kernels leads to some artifacts at the continuum edges. All
calculations are performed in the zero-temperature limit, $T=0$.

Unless otherwise specified, the model parameters are $U/D=1$,
$\Delta/U=0.02$, $\epsilon_d=-U/2$.

\begin{figure}[t]
\centering
\includegraphics[clip,width=0.5\textwidth]{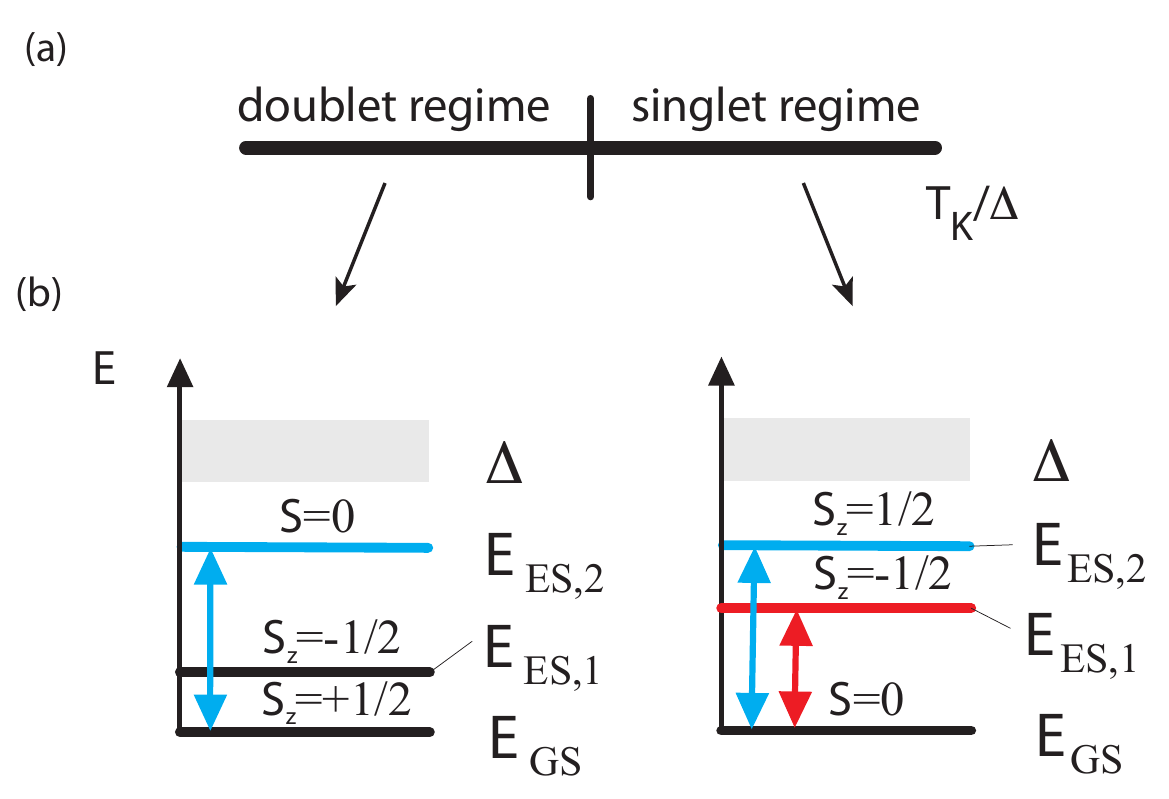}
\caption{(a) Schematic phase diagram for $B=0$. (b) Sub-gap splitting
for finite field $B$.}
\label{fx}
\end{figure}

The ground state of the Anderson impurity model, Eq.~(\ref{model}), in
the absence of the magnetic field is either a singlet or a doublet
depending on the ratio of the Kondo temperature
\cite{wilson1975,krishna1980a} $T_K \approx 0.18 U \sqrt{8\Gamma/\pi
U} \exp(-\pi U/8\Gamma)$ and the superconducting gap $\Delta$. The
impurity spin is screened by the conduction electrons for $\Delta <
\Delta_c$ forming a spin singlet, while for $\Delta > \Delta_c$ the
local moment is unscreened and the ground state forms a spin doublet;
here $\Delta_c \approx T_K/0.3$
\cite{satori1992,sakai1993,yoshioka2000} in the limit $U/\Gamma \ll
1$. At the quantum phase transition the energy of the excited
many-particle state goes to zero, and the energy levels cross. The
transition is accompanied by a jump in the spectral weight of the
in-gap resonances and a change of sign of the pairing amplitude
\cite{choi2004josephson}. The Zeeman field $B$ lifts the degeneracy of
the doublet state \cite{aniso,lee2014,zitko2015shiba,wentzell2016}.
For a spin singlet ground state, the in-gap resonances corresponding
to the doublet state are split in the magnetic field $B$. In the case
of doublet ground state, the positions of the singlet Shiba resonances
are shifted in the Zeeman field.

\begin{figure*}[t]
\centering
\includegraphics[clip,width=1.0\textwidth]{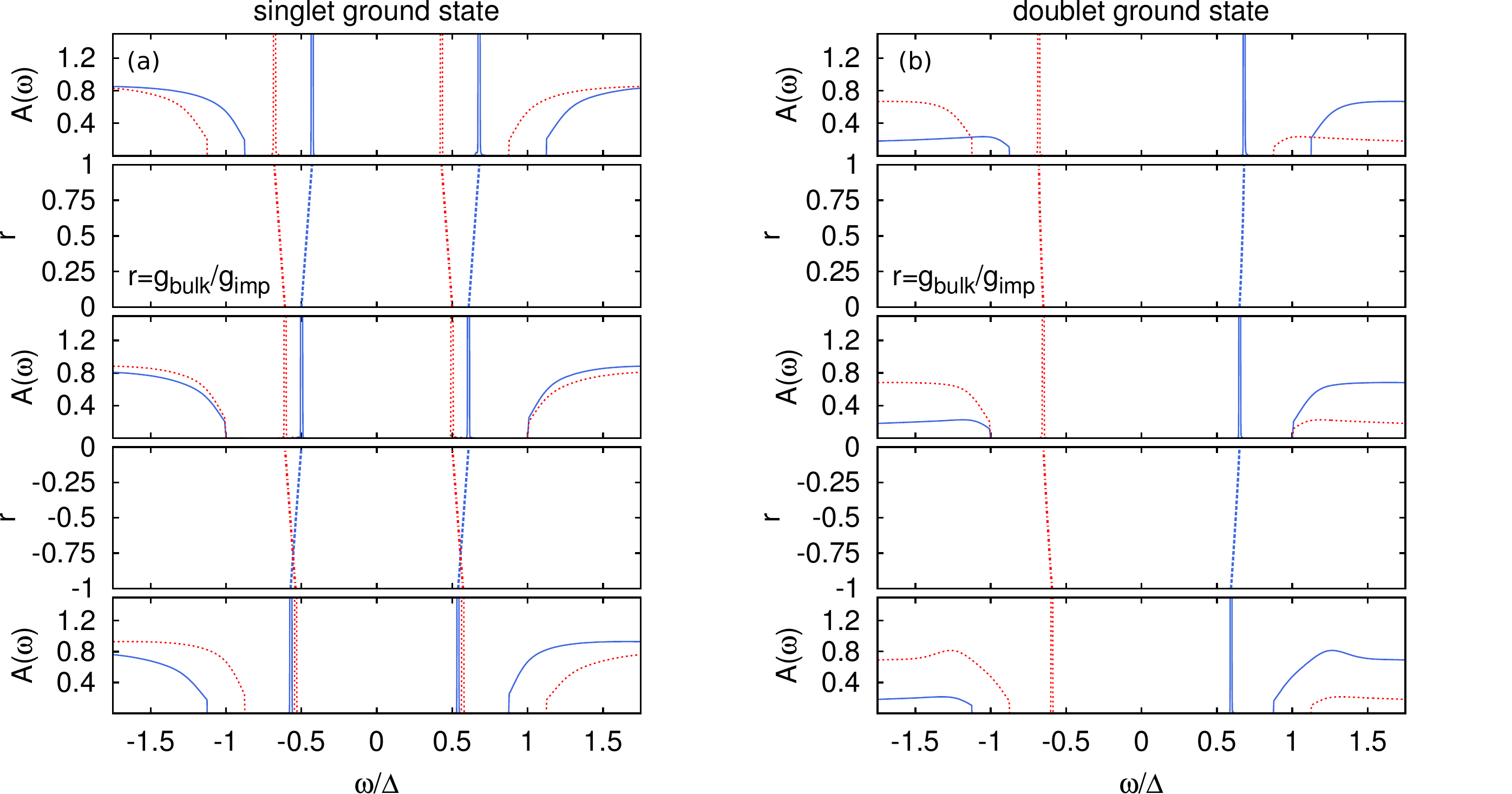}
\caption{Spectral function of the impurity for the spin singlet (a) and spin
doublet ground state (b). The parameters are $b_\mathrm{{imp}}/U=0.005$,
$\Delta/U=0.02$. For the singlet ground state $\Gamma/U=0.2$ and for the
doublet $\Gamma/U=0.075$. The spectrum for $r=\gbk/\gimp=0$ is shown in
central panels, the adjoining panels show the evolution of the position of
the Shiba resonances as $|r|$ increases, and the top/bottom panels correspond to
$r=1$ and $r=-1$, respectively.
}
\label{quantum_resonances}
\end{figure*}

Fig.~\ref{fx} shows a schematic phase diagram in zero magnetic field
and the evolution of the energy levels of the ground and excited
states with increasing Zeeman magnetic field. This evolution of the
in-gap resonances with changes of the hybridization and the magnetic
field has been recently observed in tunneling experiments and agrees
with the theoretical predictions in the case when the field is coupled
only with impurity
\cite{lee2014,jellinggaard2016,zitko2015shiba,wentzell2016}. Here, we
explore the fate of the subgap states when the magnetic field is also
Zeeman coupled with the bulk superconductor.

\subsection{Results}

We now discuss the spectral function of the impurity in different
parameter regimes and identify the boundary of the singlet-doublet
phase transition in the $(B,\Delta)$ parameter plane for different
values of the $g$-factor ratio $r$. 

\begin{figure}
\centering
\includegraphics[clip,width=0.4\textwidth]{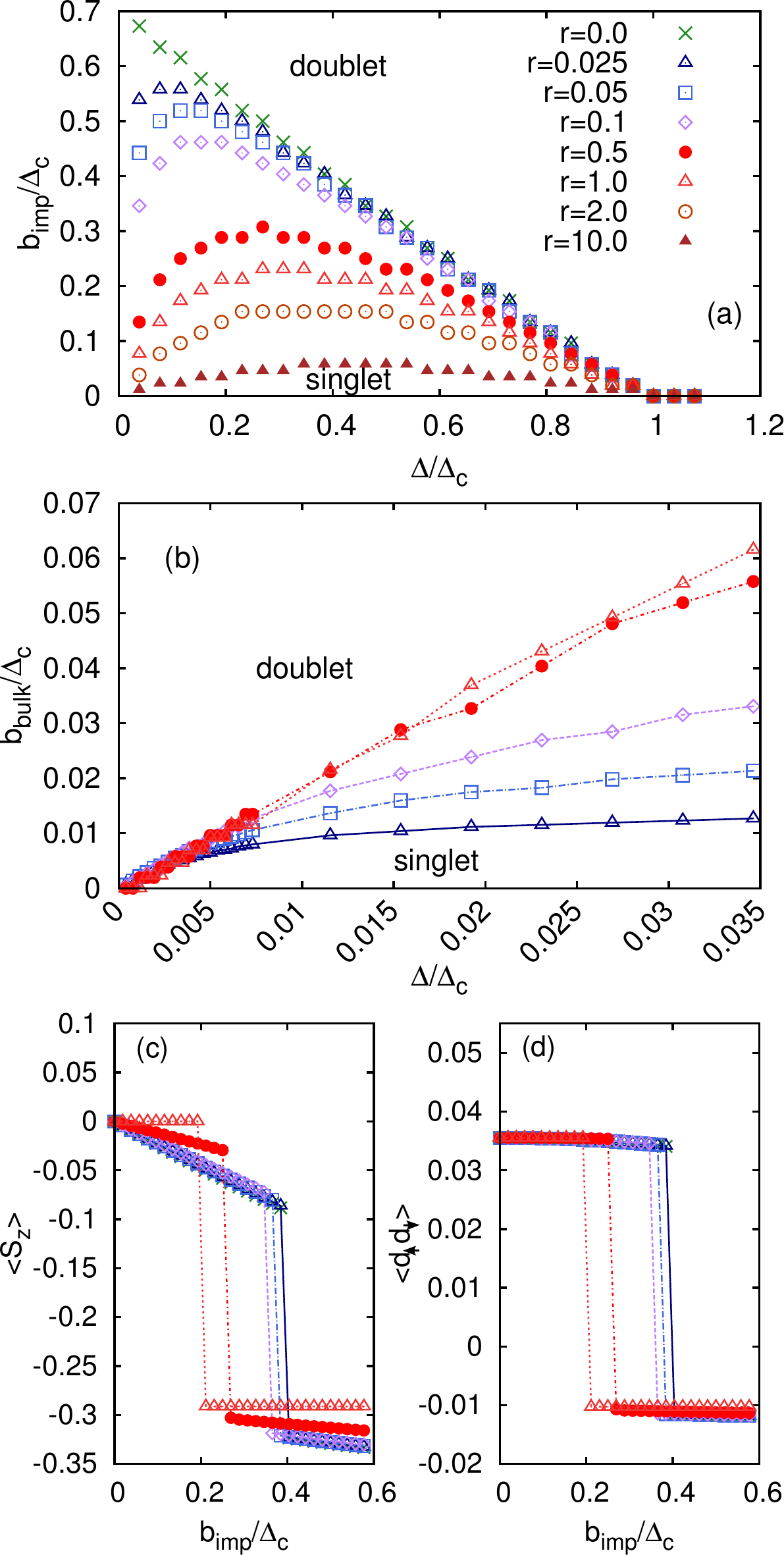}
\caption{(a) Phase diagram in the $(B,\Delta)$ plane for several
values of $r=\gbk/\gimp$. Here $\Gamma/U=0.2$, $\Delta_c/U \approx 0.13$. 
(b) For small $\Delta$ the singlet-doublet
transition coincides with the closure of the SC gap for
$b_\mathrm{bulk} \approx 2 \Delta$. (c) The expectation value
$\langle S_z \rangle$ and (d) the pairing amplitude 
$\langle d_{\uparrow} d_{\downarrow} \rangle$
abruptly change across the phase transition. Here $\Delta = 0.385 \Delta_c$.}
\label{Fig_phase_diagram}
\end{figure}

We first consider the case of singlet ground state. 
%For $B=0$, the subgap resonance is at $E=...$. 
In the magnetic field the subgap resonance (which is a spin doublet)
splits to its spin up and spin down components. The impurity spectral
function for $\Gamma/U=0.2$, $b_{\mathrm{imp}}/U=0.005$ is shown in
Fig.~\ref{quantum_resonances}(a) for $r=0$ (central panel), $r=1$ (top panel), and
$r=-1$ (bottom panel). The additional panels show how the position of
the resonances shifts as the parameter $r$ is varied. For $r=1$ the
expectation value of the spin projection $\langle S_z \rangle$ at the
impurity site is $\langle S_z \rangle = 0$ (see
Fig.~\ref{Fig_phase_diagram}(c) and Appendix A). Such compensation
holds also in the particle-hole asymmetric case as long as
$\gimp=\gbk$. If the $g$-factors are different, there will be net
magnetization at the impurity site even if the ground state is a spin
singlet and there is a finite gap to excited states.

We next consider the case of smaller hybridization, $\Gamma/U=0.075$,
so that the impurity is in the doublet ground state. The spectral
functions for $r=0$, $r=1$ and $r=-1$, as well as the evolution
between them, are shown in Fig.~\ref{quantum_resonances}(b). A single resonance is
now visible for $\omega > 0$, since the ground state has spin
projection $S_z = −1/2$, and the only possible excitation is adding a
spin-up particle to form a $S_z = 0$ singlet state. We also observe
notable differences in the appearance of the gap edges for both spin
projections, related to the strong spin polarization of the impurity
state in the doublet regime. We emphasize that this distinguishing
feature is not present in the classical impurity model discussed
above.

The phase diagram in the $(B,\Delta)$ plane is shown in
Fig.~\ref{Fig_phase_diagram}. This plot represents the main result of
this work. In the absence of a magnetic field, the ground state
changes from singlet to doublet for $\Delta = \Delta_c = 0.13 U$.
Here, $T_K \approx 0.018 U$ and $T_K/\Delta_c = 0.138$ for the chosen
value of $\Gamma/U=0.2$. For $\Delta < \Delta_c$ the transition can be
also induced by changing the magnetic field. For $r=0$ the magnetic
field is coupled only with the impurity. In this case, as shown in
Ref.~\onlinecite{zitko2015shiba}, the critical magnetic field $B_c$
for the singlet-doublet transition linearly depends on the gap, $B_c
\sim \Delta_c - \Delta$. For $r \neq 0$, however, $B_c$ has
non-monotonic dependence on $\Delta$: it increases approximately
linearly with $\Delta$ as it gets reduced from $\Delta_c$, reaches a
maximum and then decreases to zero as $\Delta \rightarrow 0$. For
$\Delta \sim \Delta_c$ the singlet-doublet transition is a consequence
of a competition of three characteristic energies: $\Delta$, $T_K$ and
$B$. For very small values of $\Delta$ (for $\Delta \ll \Delta_c$) the
singlet-doublet transition coincides with the closure of the
superconducting gap for $b_\mathrm{bulk}=2\Delta$. The phase boundary
for small value of $\Delta$ is shown in
Fig.~\ref{Fig_phase_diagram}(b). We note that for small $\Delta$ the
transition to the normal phase would actually occur for smaller value
of $B$, $B=B_{\mathrm{cl}}=\sqrt{g}\Delta \approx \sqrt{2} \Delta$,
known as the Clogston limit \cite{clogston1962,chandrasekhar1962}. For
$B > B_{\mathrm{cl}}$ the normal phase has lower free energy than the
superconducting one. Our main focus is, however, on larger values of
the superconducting gap when it is comparable to the Kondo
temperature.

The average value of the projection of the local spin $\langle S_z
\rangle$ abruptly changes at the phase transition,
Fig.~\ref{Fig_phase_diagram}(c).  For
$g_{\mathrm{imp}}=g_{\mathrm{bulk}}$, i.e., for $r=1$, the average
value $\langle S_z \rangle =0$ in the singlet case (see also Appendix
A). For $r \neq 1$, $\langle S_z \rangle$ is nonzero but small for
singlet ground state, and it jumps to large absolute value by
increasing the magnetic field at the transition to doublet ground
state.  The pairing amplitude on the impurity, $\langle d_{\uparrow}
d_{\downarrow} \rangle$, shows a characteristic sign change at the
transition, Fig.~\ref{Fig_phase_diagram}(d).

\begin{figure}[t]
\centering
\includegraphics[clip,width=0.4\textwidth]{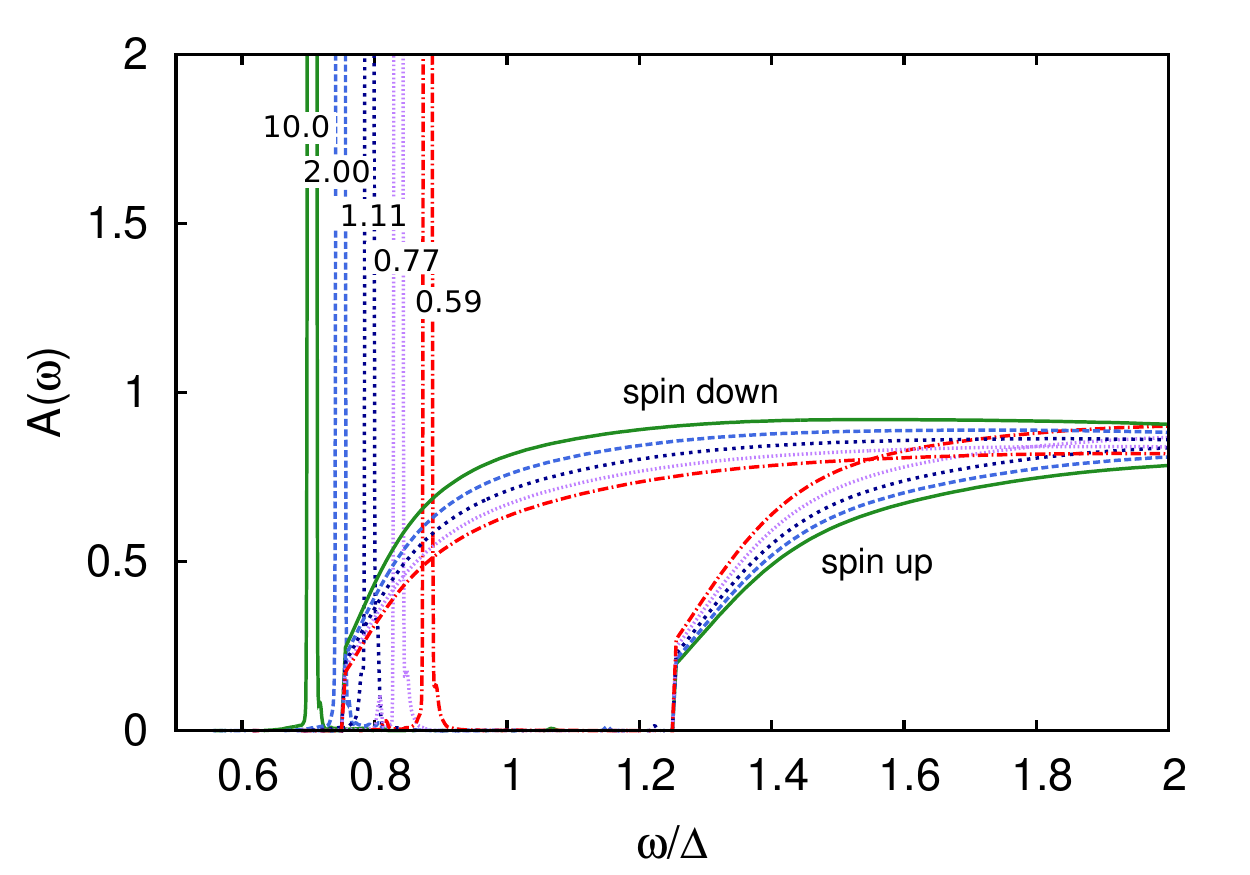}
\caption{
Spin up in-gap resonances and
continuum of excitations for several values of $r$. Here
$\bbk/U=0.01$ was kept constant. The finite width of the
Shiba resonances is a broadening artifact: these resonances are true
$\delta$-peaks at zero temperature.
}
\label{Fig_overlap}
\end{figure}

\begin{figure}
\centering
\includegraphics[clip,width=0.4\textwidth]{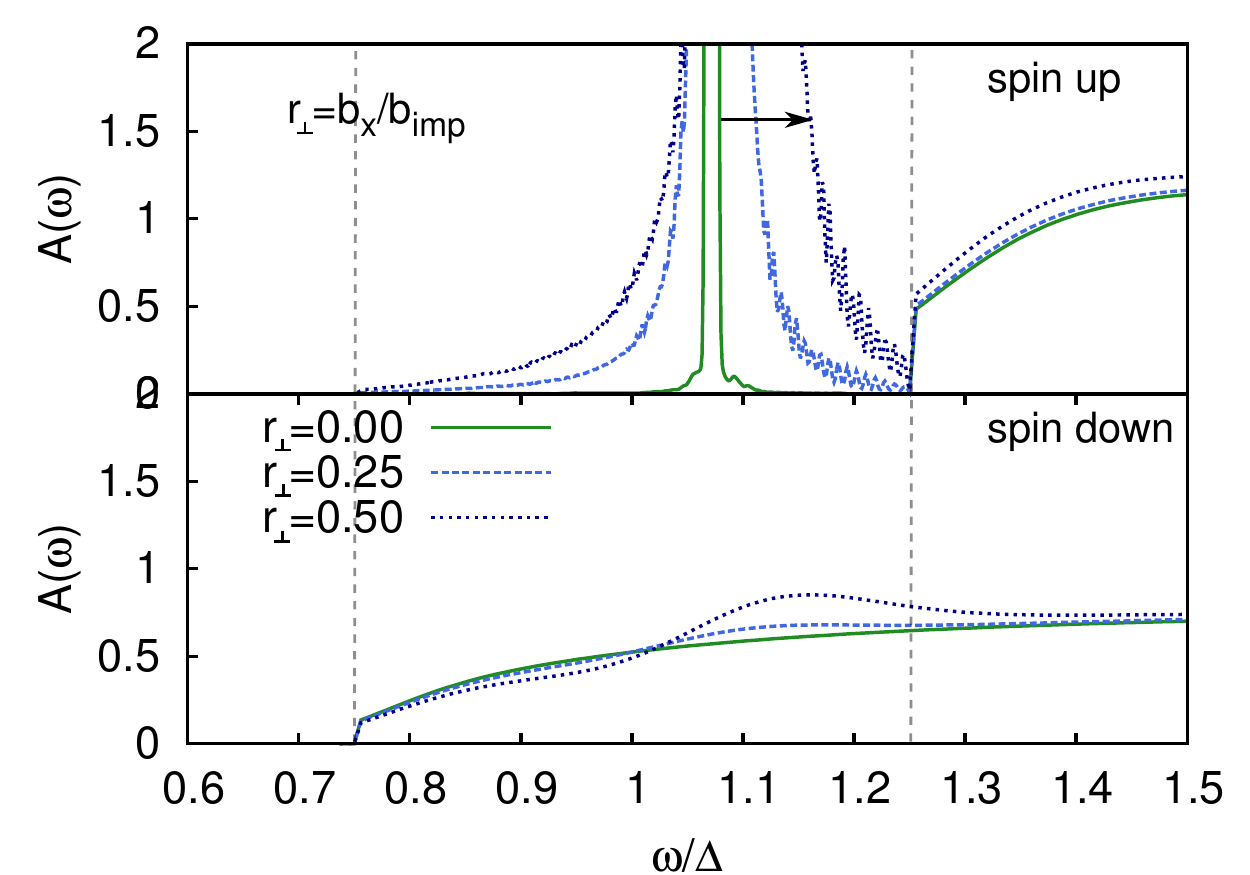}
\caption{ Spectral function of the spin up Shiba resonance and the
quasiparticle continuum for several values of the spin flipping
transverse magnetic field. As $b_x$ increases, the Shiba resonance
broadens.
}
\label{Fig_broadennig}
\end{figure}

When the spin-up Shiba state begins to overlap with the spin-down
branch of Bogoliubov excitations, it remains unperturbed, as in the
classical impurity model. This is the case in spite of the spin-flip
processes in the quantum model, and is a simple consequence of the 
conservation of the spin projection $S_z$ quantum number. In other
words, the spin-up Shiba state is a bound state of spin-up Bogoliubov
quasiparticles which are orthogonal to and do not mix with the
spin-down Bogoliubov quasiparticles. This is illustrated in
Fig.~\ref{Fig_overlap}. Here $\gbk$ and $B$ were kept constant, while
the position of the up-spin resonance was changed by changing $\gimp$.
A transverse magnetic field, however, flips the spin and the
Shiba resonances broaden, as illustrated in Fig.~\ref{Fig_broadennig}.
Such broadening effects are expected in realistic systems due to SO
coupling.

\label{broad}

\section{Conclusion}

We have analyzed the behavior of magnetic impurities coupled to
superconductors subject to an applied magnetic field that does not
fully suppress the superconducting order but strongly spin-splits the
Bogoliubov quasiparticle continua because of the Zeeman coupling. This
situation commonly occurs when the field is applied in the plane of a
superconducting thin layer and leads to clearly observable effects.

For a classical impurity, approximated as a static local point-like
magnetic field (and aligned with the external field), we find that the
position of the Shiba state is shifted linearly with the external
field as a simple consequence of the shifting edges of the
quasiparticle bands. In fact, the only effect of the spin-splitting of
the Bogoliubov states is that the frequency argument in the impurity
Green's function is shifted as $\omega \to \omega + \bbk/2$ for spin-up
and $\omega \to \omega - \bbk/2$ for spin-down particles. The
parity-changing quantum phase transition no longer occurs at
$\alpha=\pi \rho JS/2=1$, but rather when the condition
$\bbk=g_\mathrm{bulk} \mu_B B/2 = \Delta (1-\alpha^2)/(1+\alpha^2)$ is
met. This occurs for $\alpha=\alpha^*<1$. We observed that for large
$\alpha$ the Shiba state of a given spin may overlap with the
quasiparticle continuum of the opposite spin and still remain a sharp
resonance (a $\delta$ peak). This remains true as long as there is no
matrix element linking the quasiparticles of both spins.

We then turned to the case of a quantum impurity with far more complex
behavior. The Zeeman coupling is present both in the bulk and on the
impurity site, and generically the corresponding $g$-factors are
different: this is typically indeed the case in the nanoscale hybrid
superconductor-semiconductor devices. We find a very significant
effect of the Zeeman splitting of the quasiparticle continua: the
phase diagram of the possible many-particle ground states (singlet or
doublet) in the $(\Delta,B)$ plane actually has two very different
regimes. In the $\Delta \to \Delta_c$ limiting regime, the transition
occurs because a strong enough field decreases the energy of spin-down
doublet state below that of the singlet state. In this regime, the
phase boundary in the $(\Delta,B)$ plane has a negative slope: the
closer $\Delta$ is to $\Delta_c$, the smaller the separation between
the singlet and doublet states in the absence of the field, hence a
smaller Zeeman splitting is necessary to induce the transition. We
have established that for finite $r=g_\mathrm{bulk}/g_\mathrm{imp}$
the splitting between the doublet sub-gap states is larger than for
$r=0$, hence the separation between the singlet and the spin-down
doublet is smaller, thus the transition occurs for a smaller value of
the magnetic field. In the other limiting regime of small $\Delta$,
the transition occurs because the gap between the spin-polarized
Bogoliubov bands closes and the transition line is given
asymptotically as $\bbk/2=\Delta$, hence the transition line has a
positive slope. In reality, such transition is of course preempted by
a bulk transition to the normal state (Clogston limit). Nevertheless,
even in the physically accessible regime we observe that the actual
behavior is determined by a competition of both trends and that the
slope of the transition line changes at some intermediate point where
the system crosses over from one limiting behavior to another. The
actual transition line is therefore bell-shaped and depends on the
value of $r$. The straight line found in the limit $r \to 0$ is, in
fact, highly anomalous, and for realistic values of the ratio $r$
there will be a significant degree of curvature.

We have confirmed the possibility of a sharp Shiba resonance
overlapping with the continuum of opposite-spin Bogoliubov
quasiparticles. In addition, we have considered the gradual widening
of the Shiba resonance if local spin-flip processes are allowed
(generated, e.g., by SO coupling leading to non-collinear
effective magnetic fields): such processes lead to the hybridization
of the Shiba state and its gradual engulfing in the continuum.

In conclusion, we have established the importance of including the
Zeeman splitting in the bulk of the superconductor when discussing the
effect of the external magnetic field on the sub-gap states induced by
magnetic impurities in superconductors.

\begin{acknowledgments}
W.V.vG. and D.T. were supported by the Serbian Ministry of Education,
Science and Technological Development under Project ON171017, and by
the European Commission under H2020 project VI-SEEM, Grant No.~675121.
R.~\v{Z}. acknowledges the support of the Slovenian Research Agency
(ARRS) under Program P1-0044 and J1-7259. The authors acknowledge
support from the bilateral Slovenian-Serbian project ``Strong
electronic correlations and superconductivity``.
\end{acknowledgments}

\appendix

\section{Non-interacting model}

For completeness, in this appendix we define the analytical expression
for the non-interacting Anderson impurity model ($U=0$), see also Ref. \onlinecite{machida1972bound}. We work in
the Nambu space, $D^\dag=( d^\dag_\uparrow, d_\downarrow)$, $C^\dag_k = (
c^\dag_{k\uparrow}, c_{-k\downarrow} )$. The Hamiltonian can be
written as
\begin{equation}
H_{SC} = \sum_k C^\dag_k A_k C_k,
\end{equation}
where
\begin{equation}
A_k = \begin{pmatrix}
\epsilon_k + b_\mathrm{bulk}/2 & -\Delta \\
-\Delta & -\epsilon_k + b_\mathrm{bulk}/2
\end{pmatrix}.
\end{equation}
The Green's function is given by $g_k(z) = (z - A_k)^{-1}$,
\begin{equation}
g_k(z)^{-1} = (z - b_\mathrm{bulk}/2) \sigma_0 - \epsilon_k \sigma_3 +
\Delta \sigma_1,
\end{equation}
with $\sigma_{1,2,3}$ being Pauli matrices and $\sigma_0$ the
identity matrix, so that
\begin{equation}
g_k(z) = \frac{(z-b_\mathrm{bulk}/2) \sigma_0 + \epsilon_k \sigma_3 -
\Delta \sigma_1}{(z-b_\mathrm{bulk}/2)^2-(\epsilon_k^2+\Delta^2)}.
\end{equation}
The impurity Green's function is
\begin{equation}
G(z)^{-1}(z) = z \sigma_0 - \epsilon_d \sigma_3 - (b_\mathrm{imp}/2)
\sigma_0 - V^2 \sigma_3 \frac{1}{N} \sum_k g_k(z) \sigma_3.
\end{equation}
In the wide-band limit
\begin{equation}
-V^2 \frac{1}{N} \sigma_3 \sum_k g_k(z) \sigma_3 =
\Gamma \frac{(z-b_\mathrm{bulk}/2)\sigma_0 + \Delta
\sigma_1}{E(z-b_\mathrm{bulk}/2)},
\end{equation}
where $\Gamma=\pi \rho_0 V^2$. $T \to 0$, on real axis, $z=x+i\delta$:
\begin{equation}
\begin{split}
E(x) &= -i \sgn(x) \sqrt{x^2-\Delta^2},\quad \text{ for } |x|>\Delta, \\
E(x) &= \sqrt{\Delta^2-x^2},\quad \text{ for } |x|<\Delta.
\end{split}
\end{equation}
Finally, we have
\begin{equation}
G^{-1}(\omega) = (\omega-b_\mathrm{imp}/2) \sigma_0 - \epsilon_d
\sigma_3 + \Gamma \frac{(\omega-b_\mathrm{bulk}/2)\sigma_0 + \Delta
\sigma_1}{E(\omega-b_\mathrm{bulk}/2)}.
\end{equation}
\begin{widetext}
Matrix inversion yields
\begin{equation}
G(\omega) = \frac{1}{D(\omega)}
\left[
(\omega-b_\mathrm{imp}/2)
\left( 1 + \frac{\Gamma}{E(\omega-b_\mathrm{bulk}/2)} \right)
\sigma_0
-
\frac{\Gamma \Delta}{E(\omega-b_\mathrm{bulk}/2)} \sigma_1 +
\epsilon_d \sigma_3
\right],
\end{equation}
with
\begin{equation}
D(\omega) = (\omega-b_\mathrm{imp}/2)^2 \left[
1+ \frac{\Gamma}{E(\omega-b_\mathrm{bulk}/2)} \right]^2 -
\frac{\Gamma^2 \Delta^2}{E(\omega-b_\mathrm{bulk}/2)^2}-\epsilon_d^2.
\end{equation}
\end{widetext}

Now assume $b \equiv b_\mathrm{imp} = b_\mathrm{bulk}$. We consider
two functions $G_\uparrow (\omega) = G_{11}(\omega+b/2)$ and
$G_\downarrow(\omega) = -G_{22}(-\omega-b/2)^*$. Taking into account
the symmetry properties of $E(x)$, it is easily shown that $G_\uparrow
= G_\downarrow$ both inside and outside the gap. This shows that as
long as the system is in the singlet ground state, it is possible to
shift the spectral functions of spin-up and spin-down sub-systems to
make them overlap, thus their integrals over the negative energies
(occupied states) are equal, hence $\expv{S_z}=0$. This is also the
case in the interacting case. For $\bimp \neq \bbk$, $\expv{S_z}$ in
the singlet regime will be non-zero but small. In the doublet regime,
irrespective of the value of $r=\bbk/\bimp$, $\expv{S_z}$ is large.

\bibliography{spinsplit}
%\bibliographystyle{unsrt}
%\harvardparenthesis{square}

\end{document}